\documentclass[twocolumn,showpacs,superscriptaddress,nofootinbib,aps,prl,floatfix]{revtex4-2}
\usepackage{graphicx}
\usepackage{amsmath}
\usepackage[utf8]{inputenc}
\usepackage[T1]{fontenc}
\usepackage{textcomp}
\usepackage{mathptmx}
\DeclareMathAlphabet{\mathcal}{OMS}{cmsy}{m}{n}
\usepackage{xcolor}
\usepackage{hyperref}
\hypersetup{colorlinks=true,citecolor={blue},linkcolor={blue},urlcolor={blue}}
\usepackage{soul}
\usepackage{bm}
\usepackage{cleveref}

\begin{document}
\title{Photonic Berry curvature in double liquid crystal microcavities with broken inversion symmetry} 

\author{P. Kokhanchik}
\affiliation{Skolkovo Institute of Science and Technology, Bolshoy Boulevard 30, bld. 1, Moscow, 121205, Russia}

\author{H. Sigurdsson}
\affiliation{Skolkovo Institute of Science and Technology, Bolshoy Boulevard 30, bld. 1, Moscow, 121205, Russia}
\affiliation{School of Physics and Astronomy, University of Southampton, Southampton SO17 1BJ, United Kingdom}

\author{B.\,Pi\k{e}tka}
\author{J.\,Szczytko}
\affiliation{Institute of Experimental Physics, Faculty of Physics, University of Warsaw, ul.~Pasteura 5, PL-02-093 Warsaw, Poland}

\author{P.G. Lagoudakis}
\email[]{p.lagoudakis@skoltech.ru}
\affiliation{Skolkovo Institute of Science and Technology, Bolshoy Boulevard 30, bld. 1, Moscow, 121205, Russia}
\affiliation{School of Physics and Astronomy, University of Southampton, Southampton SO17 1BJ, United Kingdom}

\date{\today}

\begin{abstract}
We investigate a photonic device consisting of two coupled optical cavities possessing Rashba-Dresselhaus spin-orbit coupling, TE-TM splitting, and linear polarisation splitting that opens a tuneable energy gap at the diabolic points of the photon dispersion; giving rise to an actively addressable local Berry curvature. The proposed architecture stems from recent advancements in the design of artificial photonic gauge fields in liquid crystal cavities [K. Rechci\'{n}ska et al., Science \textbf{366}, 727 (2019)]. Our study opens new perspectives for topological photonics, room-temperature spinoptronics, and studies on the quantum geometrical structure of photonic bands in extreme settings.
\end{abstract}

\pacs{}

\maketitle 

{\it Introduction ---} Shaping and molding the optical properties in structured microscale systems (e.g., photonic crystals) has both wide and important impact in quantum optics, information transfer, light-matter interactions, and for future optolectronic and spinoptronic devices~\cite{Vahala_Nature2003}. Perhaps one of the most promising outcomes of engineering cavity photon dispersions is access to photonic analogues of electronic solid state physics where the role of the electron spin is instead played by the vectorial composition of the photon polarization. Indeed, designing artificial gauge fields for photons~\cite{Hey_RoySoc2018} has resulted in a surge of research dedicated to topological photonics~\cite{Lu_NatPho2014, Ozawa_RMP2019} and synthesis of photonic spin-orbit coupling (SOC) Hamiltonians~\cite{Leyder_NatPhy2007, Bliokh_NatPho2015, Sala_PRX2015} in a similar spirit to advancements in cold atoms~\cite{Dalibard_RMP2011} and solids~\cite{VOZMEDIANO_PhyRep2010}.

Recently, flexible liquid crystal (LC) microcavites have displayed an amazing ability to tune their cavity photon dispersions between the TE and TM polarized modes, realizing synthetic spin-orbit coupling of light~\cite{Lekenta_LightScience2018, rechcinska2019engineering}. The flexibility stems from the voltage dependent orientation of the LC molecular director allowing one to control the dielectric tensor of the cavity by adjusting the voltage applied to the LC. Moreover, since LC cavities operate at room-temperature conditions they are highly favorable in bringing complex applied photonic architectures reliant on artificial gauge fields, and topological photonics, closer to commercial use. 

\begin{figure*}
\includegraphics[width=170mm]{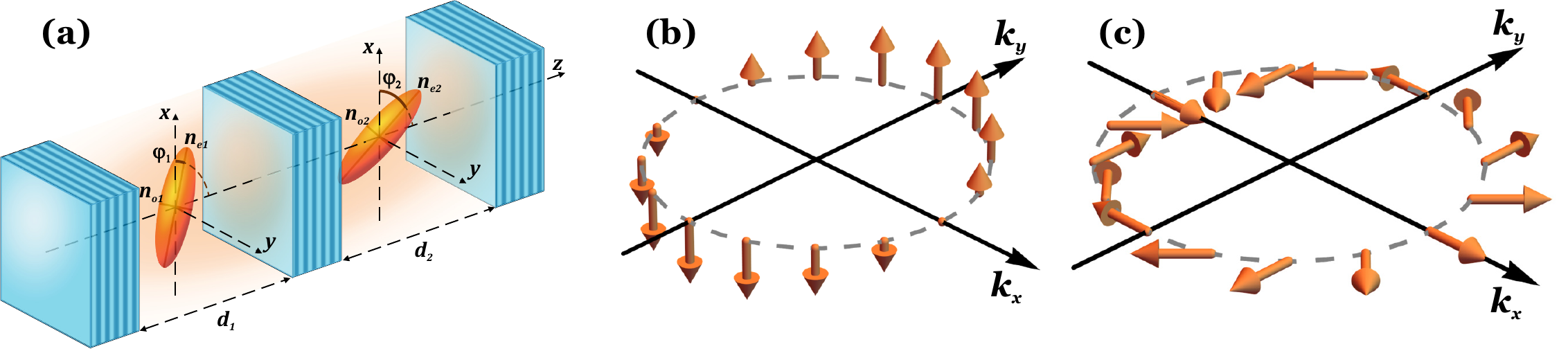}
\caption{\label{fig_1}\textbf{Scheme of the double microcavity system and the effective magnetic fields.} (a) Cavity structure showing the refractive index ellipsoid. Blue-ish stacked layers indicate distributed Bragg reflectors (DBRs). The ordinary and extraordinary refractive indices of the LC are denoted by $n_{oi}$ and $n_{ei}$ in the $i$th cavity. Panels (b) and (c) represent effective RD and OSHE induced magnetic fields (orange arrows) respectively belonging to each cavity. In (c) we don't take into account the additional static $XY$ splitting term  in Eq.~\eqref{eq.OSHE} (i.e., $\beta=0$).}
\end{figure*}

With this development, a new generation of devices can be constructed of compound cavity systems which hybridize distinct gauge fields in different cavities to produce more exotic photonic gauge fields~\cite{Umucalilar_PRA2011} (see Fig.~\ref{fig_1}a). Double microcavity systems have already been studied in both the weak coupling regime (i.e., photon lasing)~\cite{Bayer_PRL1998, Vasconcellos_APL2011, Noble_JModOp2016} and in the strong-coupling regime~\cite{Ardizzone_SciRep2013, Luk_JOptB2018, sciesiek2020long}. Of interest, it was shown in longitudinally coupled cavities~\cite{Lafont_APL2017} that exciton-polariton quasiparticles possessed excitation power-dependent spin textures appearing from interplay between their inherited photonic SOC and exciton-exciton interactions. Although anisotropic spinor polariton interactions can produce effective magnetic fields for topological purposes~\cite{Sigurdsson_PRB2019} it is unpractical in a pure photonic setting where nonlinearities are weak. On the other hand, compound LC cavity systems have not been considered until now to provide access to new photonic gauge field physics.

Another equally important advantage of combining two microcavities is the appearance of a non-trivial Berry curvature~\cite{berry1984quantal}, which plays a significant role in physics. Recently, the non-Abelian nature of a cavity photon gauge field was intimately linked to high energy physics~\cite{fieramosca2019chromodynamics} through identification of terms shared between photon spinor equation of motion and the Yang Mills model. The Berry curvature is linked with transport phenomena which includes the anomalous velocity that gives rise to a Hall current~\cite{karplus1954hall} and also the quantum Hall effect~\cite{thouless1982quantized}. Also it reforms interactions in the Berry-Fermi liquid theory~\cite{chen2017berry} and it is deeply connected to crystal polarization in solid state physics~\cite{king1993theory}. Finally, Berry curvature quantifies the properties of topological materials such as 3D~topological insulators, Weyl semimetals, Dirac materials, etc. Given the importance of the Berry curvature the ability to design a system that readily produces Berry curvature for photons at room-temperature without any external magnetic fields, photoresponsive materials, or nonlinear high-intensity effects is highly demanded.

In this work we demonstrate how a simple system of two coupled microcavities containing LCs, as shown in Fig.~\ref{fig_1}a, each possessing different SOC mechanisms give rise to a gap opening at photonic Dirac points~\cite{Bleu_PRB2017, Fang_PRL2019} with the formation of non-zero local Berry curvature~\cite{Cheng_PRL2020} which quantifies important physical properties like the Chern number and intrinsic anomalous Hall conductivity in electron systems. We also show that the given Berry curvature distribution generates anomalous Hall drift for photon wavepackets in a potential gradient with band dependent {\it Zitterbewegung}.

{\it Results ---} We consider a double microcavity like shown in Fig.~\ref{fig_1}a where both cavities are filled with a LC. The birefringence of the LC in the notation given in Fig.~\ref{fig_1} is written $ \Delta n_i = n_{ei} - n_{oi}$ where $i$ denotes the cavity in question. The effective refractive index for $x$-polarized light is written:
\begin{equation}
    n_{\text{eff},i} = \frac{n_{oi} n_{ei}}{\sqrt{n_{oi}^2 \cos^2{(\varphi_i)} + n_{ei}^2 \sin^2{(\varphi_i)} }},
\end{equation}
while for $y$-polarized light refractive index is always equal $ n_{oi} $ for the $i$th cavity. For normal incident light of wavelength $\lambda$ the resonance condition for horizontal and vertical polarized light (along the $x$ and $y$ axis respectively) can be written,
\begin{equation}
m_{xi} = \frac{2 d_i n_{\text{eff},i}}{\lambda}, \qquad m_{yi} = \frac{2 d_i n_{oi}}{\lambda},    
\end{equation}
where $d_i$ denotes the $i$th cavity size.

It was recently shown that a photonic equivalent of equal Rashba-Dresselhaus (RD) SOC can be synthesized in a single LC cavity~\cite{rechcinska2019engineering} through voltage dependent tuning of its LC director. There, cavity modes of orthogonal polarizations and opposite parity were tuned into resonance by rotating the molecular director such that their coupling resulted in an effective RD SOC for the cavity photons. Here, the LC filled cavity in our double cavity system is taken to possess a RD Hamiltonian which, in reciprocal-space representation and circular polarization basis of the cavity field $|\Psi \rangle = (\psi_+, \psi_-)^t$, is written: 
\begin{equation} \label{eq.RD}
    \hat{H}_\text{RD} = \frac{\hbar^2 k_x^2}{2M_x} + \frac{\hbar^2 k_y^2}{2M_y} - 2\alpha 
    k_y \hat{\sigma}_z .
\end{equation}
Here, $\hat{\sigma}_{x,y,z}$ are the Pauli matrices, $M_{x,y}$ is the cavity photon mass along the $x$ and $y$ planar coordinates, $k_{x,y} = k (\cos{(\varphi)}, \sin{(\varphi)})$ are the in-plane momenta, and $\alpha$ is the strength of the RD SOC. The last term in Eq.~\eqref{eq.RD} can be represented as an effective magnetic field $\mathbf{B}_\text{RD} = (0; 0; -2 \alpha k_y)$ [see Fig.~\ref{fig_1}(b)] acting on the photon pseudospin $\mathbf{S} = \langle \boldsymbol{\hat{\sigma}} \rangle$ where $\boldsymbol{\hat{\sigma}} = (\hat{\sigma_x}; \hat{\sigma_y}; \hat{\sigma_z})$ is the Pauli matrix vector.

On the other hand, the second cavity in our double LC cavity system is tuned to possess two polarization dependent mechanisms in correspondence with recent studies~\cite{Lekenta_LightScience2018, rechcinska2019engineering}. First, a splitting between the TE and TM polarized modes which results in a unique photonic SOC described by an effective in-plane magnetic field which winds itself twice around the momentum space origin, whereas, in comparison, Rashba and Dresselhaus SOCs wind only once. The TE-TM splitting results in the optical spin Hall effect (OSHE)~\cite{kavokin2005optical} and is a source of multiple interesting features relevant to topological photonics~\cite{Bleu_PRB2017, Klembt_Nature2018, gianfrate2020measurement}. Second, a static splitting term between the linearly polarized modes of the photons (referred here as a $XY$ splitting). Such cavity can be described by the Hamiltonian:
\begin{equation} \label{eq.OSHE}
    \hat{H}_\text{OSHE} = \begin{pmatrix} \dfrac{\hbar^2 k_x^2}{2M_x} + \dfrac{\hbar^2 k_y^2}{2M_y} & \beta + \gamma k^2 e^{-i2\varphi} \\
    \beta + \gamma k^2 e^{i2\varphi} & \dfrac{\hbar^2 k_x^2}{2M_x} + \dfrac{\hbar^2 k_y^2}{2M_y} \end{pmatrix}.
\end{equation}
The TE-TM and $XY$ splitting are denoted by $\gamma$ and $\beta$ respectively in Eq.~\eqref{eq.OSHE}. Note the $2\varphi$ dependence indicating the double winding of the effective in-plane magnetic field $\mathbf{B}_\text{OSHE} = (\beta + \gamma ( k_x^2 - k_y^2 ) ; 2 \gamma k_x k_y; 0)$ [see Fig.~\ref{fig_1}(c)]. We stress that the acronym ``OSHE'' is just used to differentiate from $\hat{H}_\text{RD}$ since ``TE-TM + XY'' is a bit cumbersome. The optical spin Hall effect should not be confused with spin Hall or valley Hall effects~\cite{kavokin2005optical} as we are not considering fermions undergoing impurity scattering or in external electric fields.

\begin{figure*}
\includegraphics[width=170mm]{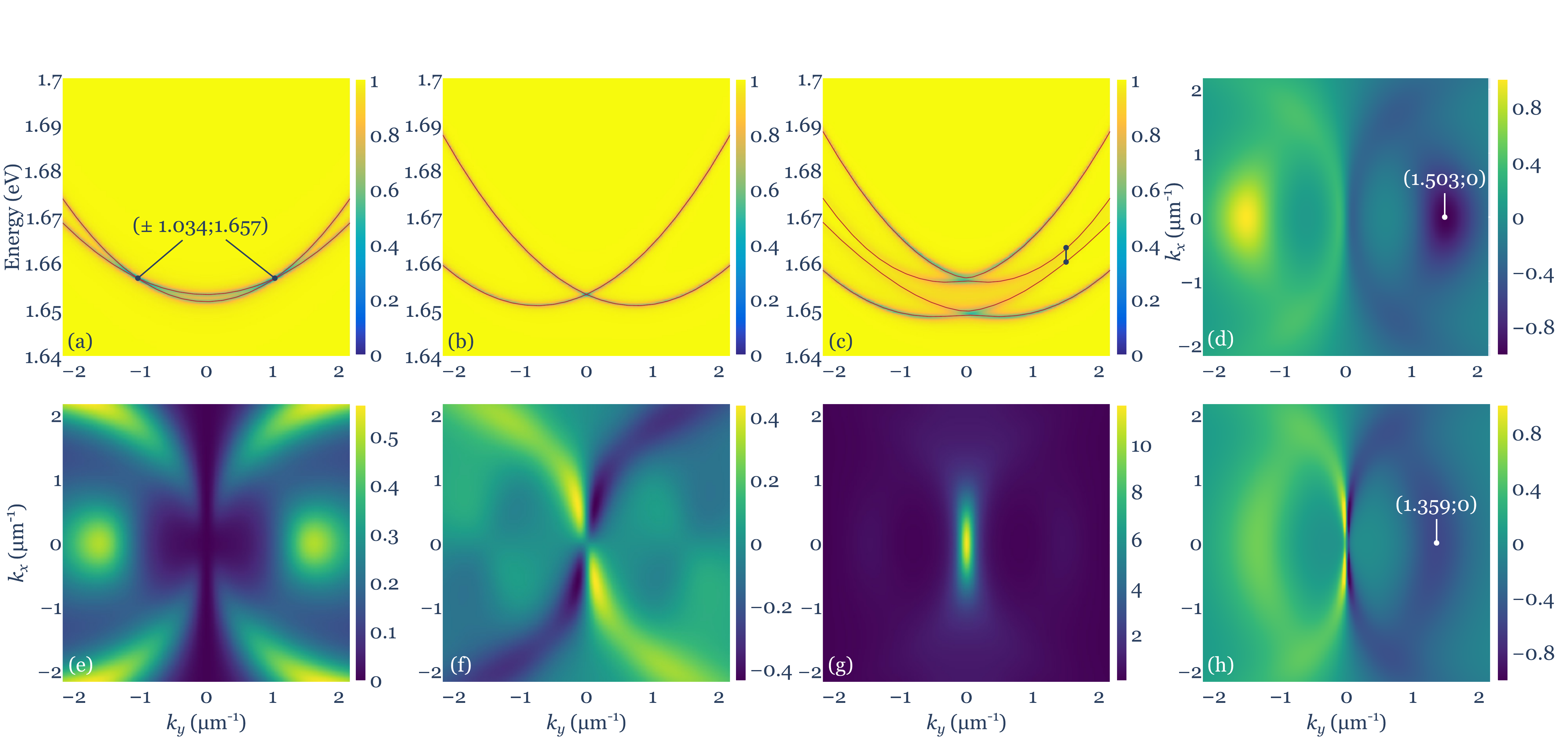}
\caption{\label{fig.bands+QGT}\textbf{Gap opening of the hybridized OSHE and RD double cavity system.} Dispersions (normalized reflection intensities using diagonally polarized incident light) of the (a) single TE-TM + XY cavity where we mark the parabolas intersection, (b) single RD cavity, and (c) both cavities coupled calculated using Berreman method~\cite{Berreman_JOpt1972}. In (a-c) we have fitted the energy bands from Eqs.~\eqref{eq.OSHE},~\eqref{eq.RD} and \eqref{eq_2}, respectively, as solid red lines with a vertical black marker in (c) indicating the smallest splitting between the central bands. (d) Berry curvature $\Omega_z^{(2)}$ calculated for the second lowest band, and the QGT components (e) $g_{xx}^{(2)}$, (f) $ g_{xy}^{(2)} = g_{yx}^{(2)}$, (g) $ g_{yy}^{(2)}$; (h) Berry curvature $\Omega_z^{(2)}$ for larger LC director angle. The colorscale in (h) is saturated to match (d). Parameters for Berreman calculations are in~\cite{parameters}}
\end{figure*}

As pointed out in recent works~\cite{bleu2018measuring, gianfrate2020measurement}, the presence of both TE-TM and $XY$ splitting in a single cavity leads to two energetically shifted parabolas with different effective masses which intersect into two tilted Dirac cones located at $(k_x,k_y) = (0,\pm\sqrt{\beta/\gamma})$, also referred as diabolical points [see Fig.~\ref{fig.bands+QGT}(a)]. Recently, this Hamiltonian was realized in single microcavity with embedded quantum wells and operating in the strong-coupling regime~\cite{gianfrate2020measurement}, as well as in polariton microcavity based on an optical birefringent 2D perovskite~\cite{polimeno2020tuning}. There, the emergent spinor polariton modes (two-band system) were subjected to an external out-of-plane magnetic field which, when combined with $\hat{H}_\text{OSHE}$, resulted in a topological gap opening at the Dirac points and formation of non-zero Berry curvature in momentum space.

The fact that each Hamiltonian given by Eq.~\eqref{eq.RD} and~\eqref{eq.OSHE} can be easily realized in a double cavity setting provides an opportunity to explore a new regime of local photonic Berry curvature. In order to realize a gap opening at the Dirac cones of $\hat{H}_\text{OSHE}$ for just cavity photons, instead of using magnetically susceptible polaritons~\cite{gianfrate2020measurement} (practically inaccessible in organic microcavities), we propose a four-band system of two coupled cavities,
\begin{equation} \label{eq_2}
    \hat{H} = \begin{pmatrix} \Delta E \hat{\sigma}_0 + \hat{H}_\text{OSHE} & J \hat{\sigma}_0 \\
    J \hat{\sigma}_0 & \hat{H}_\text{RD} \end{pmatrix}.
    \end{equation}
Here, $\Delta E$ is a detuning parameter between the OSHE and RD cavities, $\hat{\sigma}_0$ is the $2\times2$ identity operator and $J>0$ denotes coupling between the cavities. The photonic hybridization of the two subsystems $\hat{H}_\text{RD}$ and $\hat{H}_\text{OSHE}$ can then achieve a similar gap opening and finite local Berry curvature like reported in~\cite{gianfrate2020measurement} but with zero Chern number since our system is topologically trivial. Instead of an out-of-plane magnetic field that breaks time-reversal symmetry and opens a gap at the Dirac points, our system instead breaks inversion symmetry through the hybridization of the two cavities. One cavity possesses TE-TM and $XY$ splitting giving rise to Dirac cones in one subsystem, while the other has the RD SOC which, when coupled with the former, breaks inversion symmetry and opens the gap at the Dirac points with consequent emergence of non-zero local Berry curvature. Indeed, $\hat{H}$ is symmetric under time-reversal as follows from Maxwell equations whereas it is not for inversion $\mathcal{I} =\hat{\sigma}_0 \otimes \hat{\sigma}_x$,
\begin{equation} \label{eq.TRS}
    \mathcal{I} \hat{H}(-\mathbf{k}) \mathcal{I}^{-1} \neq \hat{H}(\mathbf{k}).
\end{equation} 
We note that an empty cavity (no LC) can possess TE-TM splitting when the photonic mode is shifted relative to the center of the DBRs stopband~\cite{panzarini1999cavity}, and $XY$ splitting can be implemented by creating an asymmetric microcavity~\cite{Demenev_PRB2017}. However, the experimentally reported TE-TM splitting values for an empty microcavity are of the order of tens of $\mu$eV~\cite{maragkou2011optical}, while the TE-TM splitting values in LC cavity are measured in the meV scale providing comparable scales for $\alpha,\beta$ and $\gamma$ values and allowing for an observable gap opening. The Stokes parameters $\mathbf{S}$ and field distribution of the double cavity transmitted light are detailed in Ref.~\cite{supplemental}.

In Fig.~\ref{fig.bands+QGT}(a) and~\ref{fig.bands+QGT}(b) we calculate the dispersion of each uncoupled cavity belonging Eqs.~\eqref{eq.OSHE} and~\eqref{eq.RD} respectively. The diabolic points, or Dirac cones, are marked with the black lines in Fig.~\ref{fig.bands+QGT}(a). When the two systems are coupled using Eq.~\eqref{eq_2} we observe a splitting between the two central bands around the Dirac point which obtain high concentration of Berry curvature satisfying $\Omega_z(\mathbf{k}) = -\Omega_z(-\mathbf{k})$. The odd parity of the curvature implies that the system is in a topologically trivial phase whereby integrating over $\Omega_z(\mathbf{k})$ gives a Chern number of $C=0$. Therefore, our system should not be confused with Chern insulators or Hall phases, but instead as means of generating local Berry curvature which can be applied to manipulate photon wavepacket transport properties~\cite{Wimmer_NatPhys2017}. In order to characterize the properties of the bands in the double cavity system we calculate the components of the quantum geometric tensor~\cite{wilczek1989geometric, bleu2018measuring} (QGT), whose real part contains the quantum metric (distance between eigenstates), and the imaginary part determines the Berry curvature.
\begin{gather} 
    T_{ij}^{(n)}(\mathbf{k}) = \sum_{m \neq n} \frac{\langle m_\mathbf{k} | \partial_{k_i} \hat{H} | n_\mathbf{k} \rangle \langle n_\mathbf{k} | \partial_{k_j} \hat{H} | m_\mathbf{k} \rangle}{(E_m(\mathbf{k}) - E_n(\mathbf{k}))^2},
    \label{eq_a1} \\
        g_{ij}^{(n)} = \text{Re} \left( T_{ij}^{(n)} \right), \qquad  \Omega_z^{(n)} = -2\text{Im} \left( T_{xy}^{(n)} \right),
    \label{eq_a2}
\end{gather}
where $ (i,j) = (x,y) $, $n = (1,2,3,4)$ denotes the number of the band from bottom to top in energy, and $ | n_\mathbf{k} \rangle $ and $E_n(\mathbf{k})$ are the $k$-dependent eigenstate and eigenenergy of $\hat{H}$.

The QGT components for the second lowest band of this system are shown in Figs.~\ref{fig.bands+QGT}(d-g) revealing that the strongest Berry curvature appears at the anticrossing of the two central bands of the system (marked with white line-dot). The black segment in Fig.~\ref{fig.bands+QGT}(c) shows the anticrossing, marking the opening point of the Dirac cone corresponding to the white point in Fig.~\ref{fig.bands+QGT}(d). Figure~\ref{fig.bands+QGT}(h) shows that the  Berry curvature extremum is shifted when the LC director is changed, underlining tunability coming from the LC cavitites. We note that the QGT components for a four-band system can be directly measured~\cite{bleu2018measuring}. The QGT components for the remaining three bands are given in Ref.~\cite{supplemental}.

For all effective masses we use notations $M_{x1}, M_{y1}, M_{x2}, M_{y2}$, where indices (1,2) correspond to $\hat{H}_\text{OSHE,RD}$ cavities respectively. We obtain $\beta=0.77$ meV, $\gamma=0.72$ meV $\mu$m$^2$, $M_{x1}=M_{y1}=0.054$ meV ps$^2$ $\mu$m$^{-2}$,  $M_{x2}=0.89 M_{x1}$, $M_{y2}=0.92 M_{x1}$, $\Delta E=-0.8$ meV, $\alpha=3.25$ meV $\mu$m, and $J = 3.6$ meV by fitting the numerically calculated dispersions [red lines in Figs.~\ref{fig.bands+QGT}(a-c)]. The effective mass $M_{x2}$ does not affect the gap opening, nor does it affect the position of the Dirac cones, it only affects the width of the distribution of the QGT parameters along the $k_x$-axis. It can therefore adopt a typical value obtained using the methods of Ref.~[\onlinecite{rechcinska2019engineering}]. We note that the plotted energies belonging to Eq.~\eqref{eq_2} are shifted to match the absolute energies obtained through numerics. 

For the case of all masses being equal and $k_x = 0$ and $\Delta E=0$, the analytical expression for the energies becomes:
\begin{align} \notag
     & E  = \frac{\hbar^2 k_y^2}{2M} \pm \frac{1}{\sqrt{2}} \bigg[ 2J^2  + 4\alpha^2 k_y^2 + \epsilon^2  \\
     & \pm \sqrt{\epsilon^2[\epsilon^2 + 4 J^2  -  8 \alpha^2  k_y^2] + 16 \alpha^2 k_y^2 [J^2  +  \alpha^2 k_y^2]}  \bigg]^{1/2}.
\end{align}
where $\epsilon = (\beta - \gamma k_y^2)$. If RD SOC is absent ($\alpha=0$) we obtain two pairs of bands separated in energy by $2J$. Instead of two Dirac points there are now four points located at the band crossings $k_y = \pm \sqrt{\beta/\gamma}$. When $\alpha \neq 0$ this degeneracy is lifted resulting in Berry curvature with extremum at the band anticrossing points [see Fig.~\ref{fig.bands+QGT}(d)]. The location of the extremum is shifted with respect to the original Dirac point location at $k_y = \pm \sqrt{\beta/\gamma}$ which, to the leading order, is given by the roots of the equation,
\begin{equation}
\frac{J^2  +  2\alpha^2 k_y^2}{\sqrt{J^2  +  \alpha^2 k_y^2}}  = \frac{k_y}{\alpha}\left(2\alpha^2  - \gamma (\beta - \gamma k_y^2) \right).
\end{equation}
The solution satisfies $|k_y| > \sqrt{\beta/\gamma}$ corresponding to the anticrossing point (the Berry curvature extremum) shifting to higher momentum values away from the original Dirac point.
\begin{figure}[t]
\centering
\includegraphics[width = \linewidth]{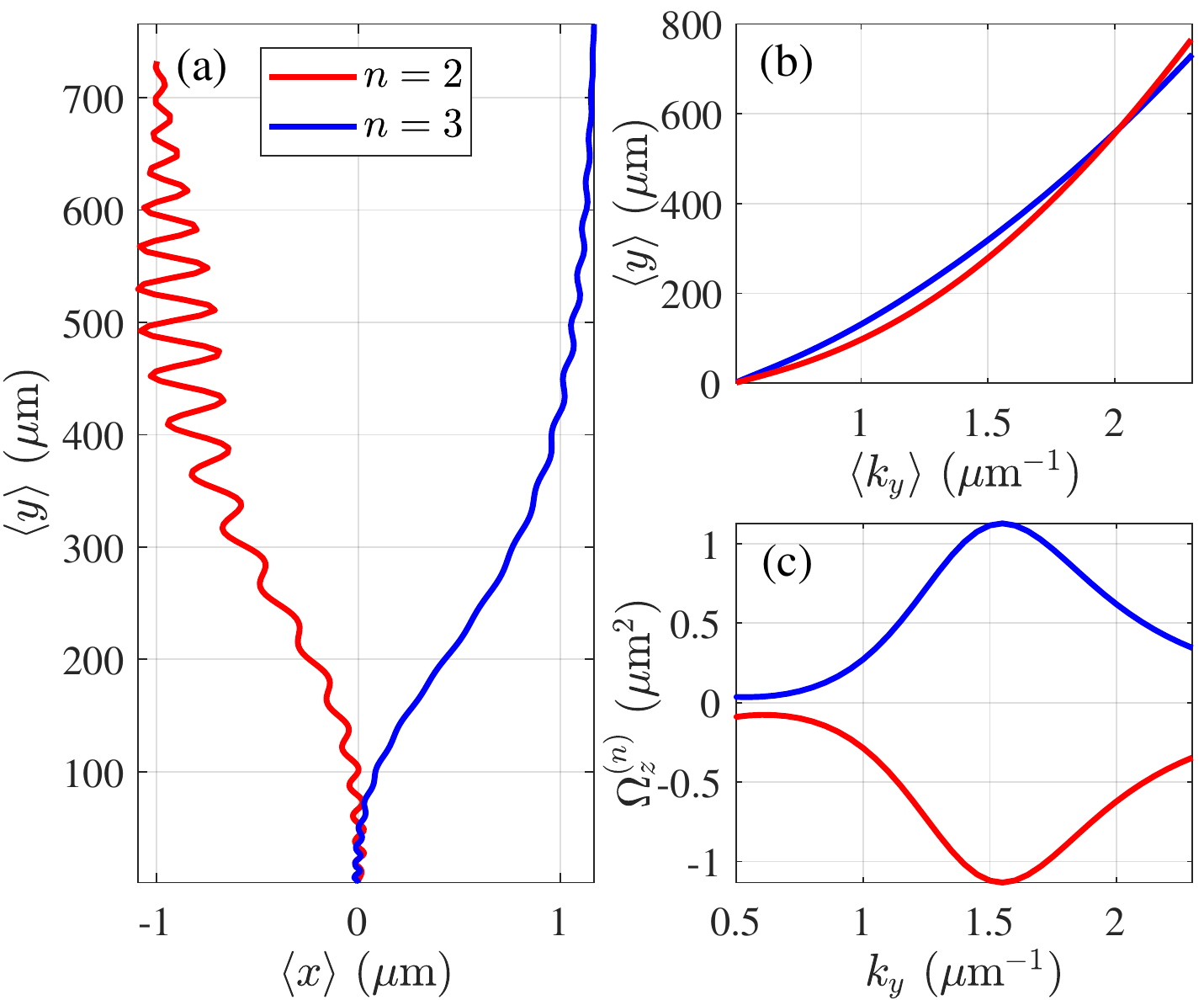}
\caption{\textbf{The anomalous Hall drift and Zitterbewegung.} (a) Photon wavepacket center of mass $\langle \mathbf{r} \rangle$ moving in a potential gradient along the $y$-direction. Solid blue and red curves correspond to a packet initialized in band $n=2$ and $n=3$ respectively. A finite drift in the $x$ direction appears due to the presence of Berry curvature $\Omega_z^{(n)}$ which possesses different signs between the bands causing the blue and red trajectories to drift in opposite directions. (b) Mean wavepacket momentum $\langle k_y \rangle$ against its vertical position and (c) the Berry curvature cross section at $k_x = 0$ of the two bands.}
\label{fig3}
\end{figure}

We will now demonstrate the anomalous Hall drift~\cite{Bliokh_NatPho2008, Wimmer_NatPhys2017} in our system appearing for an accelerated wavepacket in a band with finite Berry curvature. An accelaration in the positive $y$ direction can be provided by introducing a potential wedge in the double cavity structure which can be modeled with $V(y) = -v_0 y$ where we solve the Schrödinger equation describing the wavepacket motion. We choose two different initial conditions corresponding to bands $n = 2,3$ where a concentrated maximum of Berry curvature appears close to the point of avoided crossing [see e.g. Figs.~\ref{fig.bands+QGT}(c) and~\ref{fig.bands+QGT}(d)],
\begin{equation} \label{eq.init}
|\Psi(\mathbf{r}, t=0) \rangle = |n_{\mathbf{k}_0} \rangle e^{- r^2/2w^2} e^{i \mathbf{k}_0 \cdot \mathbf{r}}.
\end{equation}
Here, $\mathbf{k}_0 = 0.5 \ \hat{\mathbf{y}}$ $\mu$m$^{-1}$ is chosen for simplicity to avoid both the effects of strong QGT concentrations and negative effective mass at low momenta. We also set a typical value for the potential gradient at $v_0 = 45$ $\mu$eV $\mu$m$^{-1}$, and $w = 17$ $\mu$m. The wavepacket center of mass in real and reciprocal space is defined
\begin{equation}
\langle \mathbf{r} \rangle = \frac{\int \mathbf{r} \langle \Psi | \Psi \rangle \ d\mathbf{r}}{\int \langle \Psi | \Psi \rangle \ d\mathbf{r}}, \qquad  \langle \mathbf{k} \rangle = \frac{\int \mathbf{k} \langle \tilde{\Psi} | \tilde{\Psi} \rangle \ d\mathbf{k}}{\int \langle \tilde{\Psi} | \tilde{\Psi} \rangle \ d\mathbf{k}},
\end{equation}
where $\tilde{\Psi}$ is the Fourier transform of $\Psi$. Evolving Eq.~\eqref{eq.init} in time (40 ps) we observe in Fig.~\ref{fig3}(a) the two different solutions experiencing a drift in along the horizontal in the two opposite directions because $\Omega_z^{(n)}$ changes sign from band $n=2$ to $n=3$. The trajectories are not symmetric about $\langle x \rangle = 0$ because the dispersion relation of the two bands differs. Clear {\it Zitterbewegung} oscillations (which depend on the gradient and initial momentum) can be observed in the trajectories and become strong when the wavepacket moves through the anticrossing region of the dispersion. Such oscillations appear for Dirac electrons experiencing interference between their positive and negative energy states and, although quite challenging, can be artificially replicated in photonic systems~\cite{Zhang_PRL2008} like ours. In Fig.~\ref{fig3}(b) we plot $\langle k_y \rangle$ against $\langle y \rangle$ showing monotonic increase due to the potential gradient pushing the packet. The saturation of the drift in Fig.~\ref{fig3}(a) corresponds to the wavepacket having passed through the strong Berry curvature region $\Omega_z^{(n)}$ whose cross-section is shown in Fig.~\ref{fig3}(c). These anomalous optical transport effects can be tuned readily by changing the LC molecular angle by applying voltage across the cavity.

The experimental complexity of the system is primarily reduced to the task of manufacturing double cavity structure. Cavities filled with LCs should be thin enough to allow for a large separation between adjacent modes. At the same time, the contacts and internal DBR also need to be thin to implement finite observable coupling between different cavities. In principle this is only a technical challenge that could be solved by exploiting a thin polymer DBR or exfoliated DBR~\cite{rupprecht2020micro} suspended in a LC microcavity.

{\it Discussion ---} 
We have demonstrated a purely photonic implementation of achieving measurable local Berry curvature by construction of a Hamiltonian describing two optical cavities possessing distinct SOC mechanisms. These SOC mechanisms can today be readily designed through the recent advancements in tunable LC microcavities~\cite{Lekenta_LightScience2018, rechcinska2019engineering}. In our proposal, one cavity is composed to a unique photonic SOC effect stemming from both TE-TM and $XY$ splitting of the cavity modes leading to a pair of tilted Dirac cones. The second cavity provides an RD SOC which, when coupled with the former cavity, breaks inversion symmetry and leads to gap opening at the Dirac points. The opening is associated with the formation of local Berry curvature which can generate an anomalous Hall drift. Our results open new possibilites for measuring fundamental geometrical properties of photonic bands which are of great interest to the growing field of topological photonics~\cite{Ozawa_RMP2019}. The Berry curvature in our system is not reliant on gyromagnetic materials, polariton susceptibility to external magnetic fields, or complicated fabrication of transverse cavity structures like photonic honeycomb lattices~\cite{Sala_PRX2015, Cheng_PRL2020}.

{\it Acknowledgements ---} P.K. (modelling and analysis of results) and P.G.L. (led the project) acknowledge the support of RFBR Grant No. 20-02-00919. H.S. (development of model and analysis) and P.G.L. (discussion of results) acknowledge the support of UK's Engineering and Physical Sciences Research Council (Grant No. EP/M025330/1 on Hybrid Polaritonics) and European Union’s Horizon 2020 program, through a FET Open research and innovation action under the grant agreement No 899141 (PoLLoC). H.S. is additionally grateful for the hospitality of the University of Iceland. J.Z. and B.P. (model review and discussions) acknowledge the support of the National Science Centre, Poland Grants No. UMO-2019/35/B/ST3/04147 and UMO-2017/27/B/ST3/00271.

\providecommand{\noopsort}[1]{}\providecommand{\singleletter}[1]{#1}%

\setcounter{equation}{0}
\setcounter{figure}{0}
\setcounter{section}{0}
\renewcommand{\theequation}{S\arabic{equation}}
\renewcommand{\thefigure}{S\arabic{figure}}
\renewcommand{\thesection}{S\arabic{section}}
\onecolumngrid

\vspace{1cm}
\begin{center}
\Large \textbf{Supplemental Material}
\end{center}

\section{Transmitted Stokes parameters and field distribution from the double cavity}
\begin{figure}
\includegraphics[width=120mm]{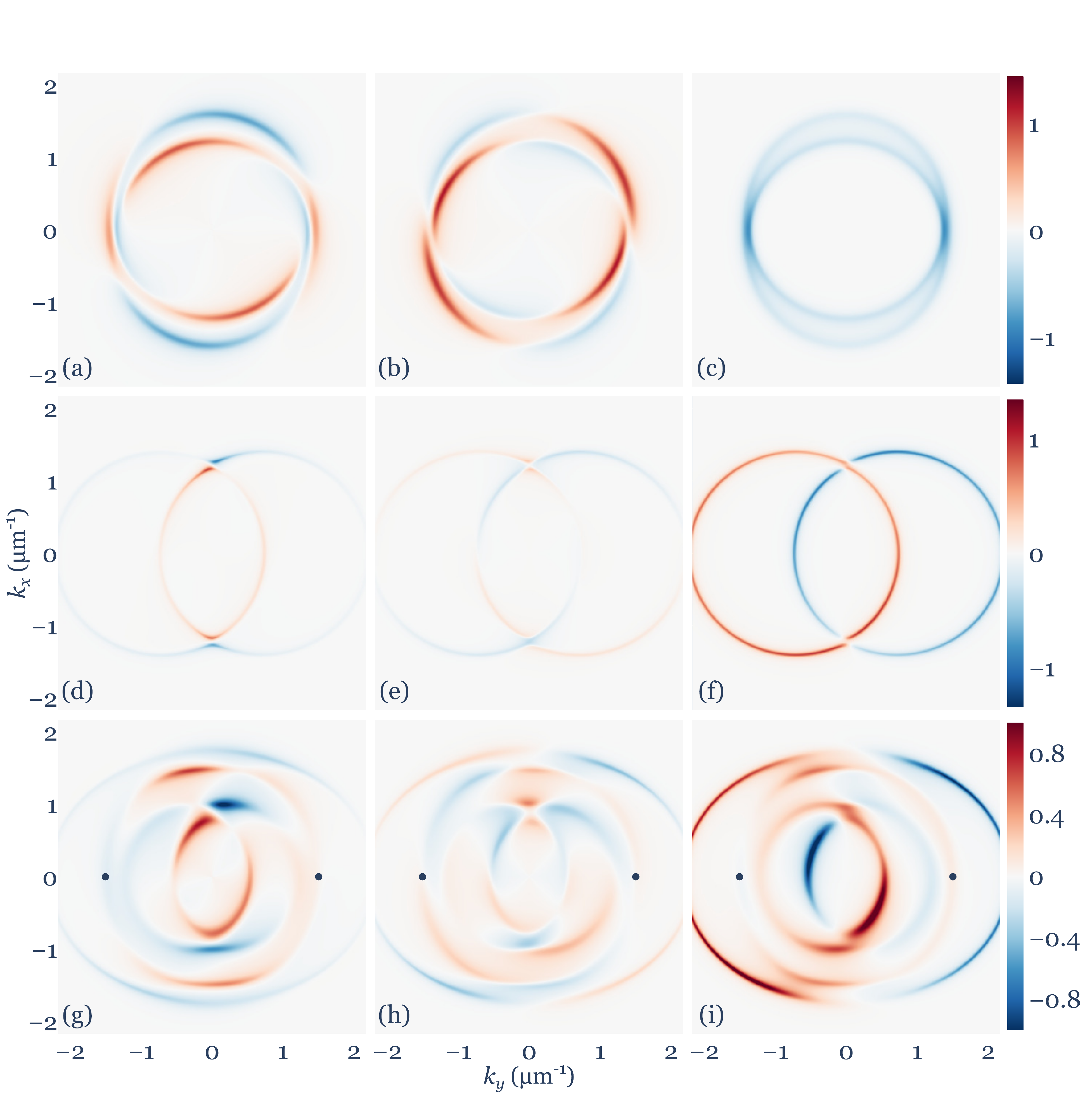}
\caption{\label{fig_s1}\textbf{Stokes parameters of the transmitted cavity light in reciprocal space.} The incident light is diagonally polarized. Calculations were performed using the Berreman method. Panels from left to right correspond to $S_1$, $S_2$ and $S_3$ respectively. Top row shows the isolated TE-TM + XY system, the middle row shows the isolated RD system, and the bottom row shows the two coupled systems. The energy $E=1.6603$ eV is the same for all figures and is chosen resonant with the second lowest band point of maximum Berry curvature. Black dots in (g,h,i) shows the position of the Berry curvature extremum.}
\end{figure}

\begin{figure}
\includegraphics[width=100mm]{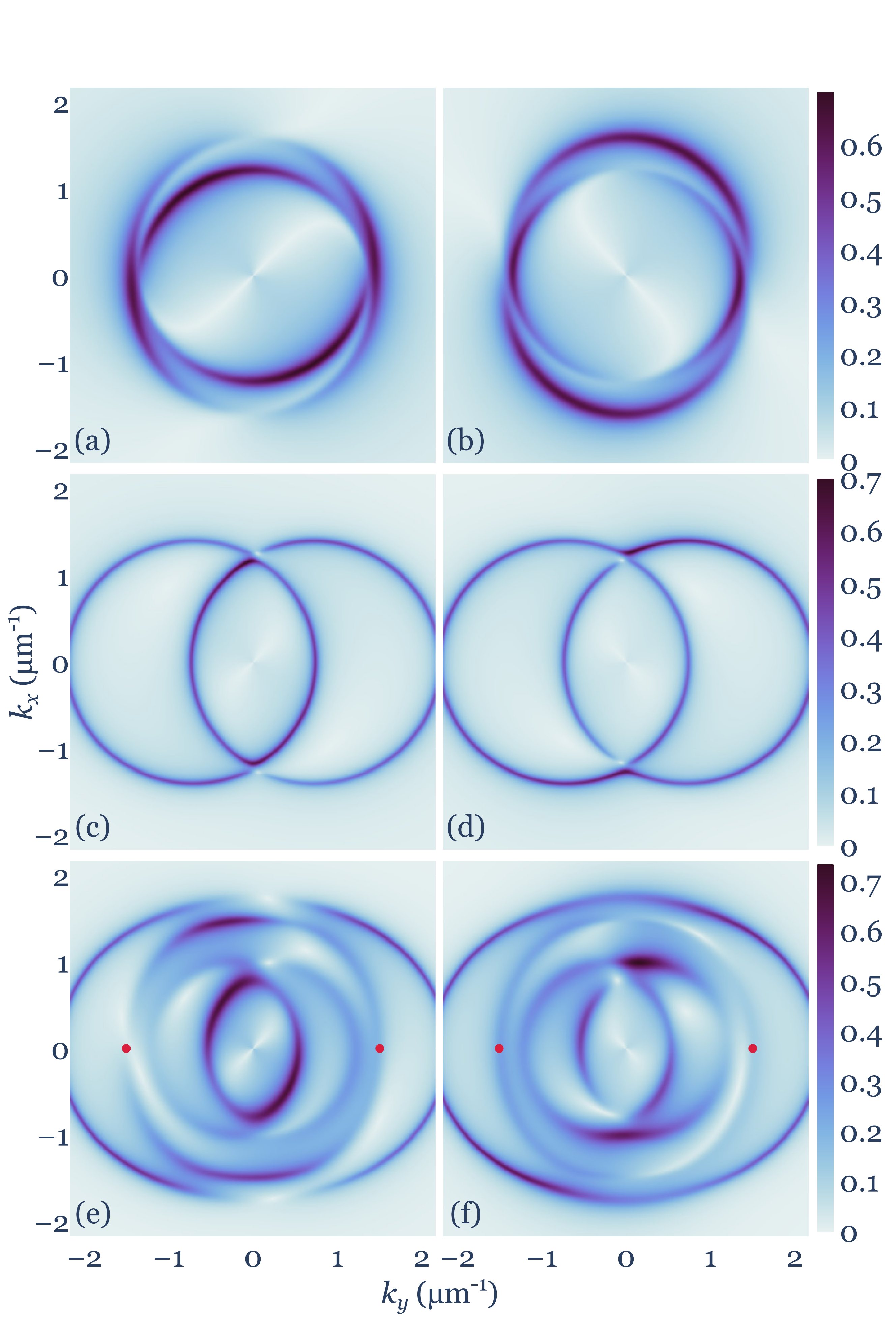}
\caption{\label{fig_s2}\textbf{Amplitudes of the transmitted electrical field components along $x$ ($|\psi_x|$, left column) and $y$ ($|\psi_y|$, right column) axes normalized to the incident light electrical field amplitude.} Calculations were performed using the Berreman method. Analogous to the order in Fig.~\ref{fig_s1} we show the field amplitudes for (a,b) isolated TE-TM + XY system, (c,d) isolated RD system, (e,f) coupled systems. Red dots in (e,f) shows the position of the Berry curvature extremum.}
\end{figure}

\begin{figure}
\includegraphics[width=120mm]{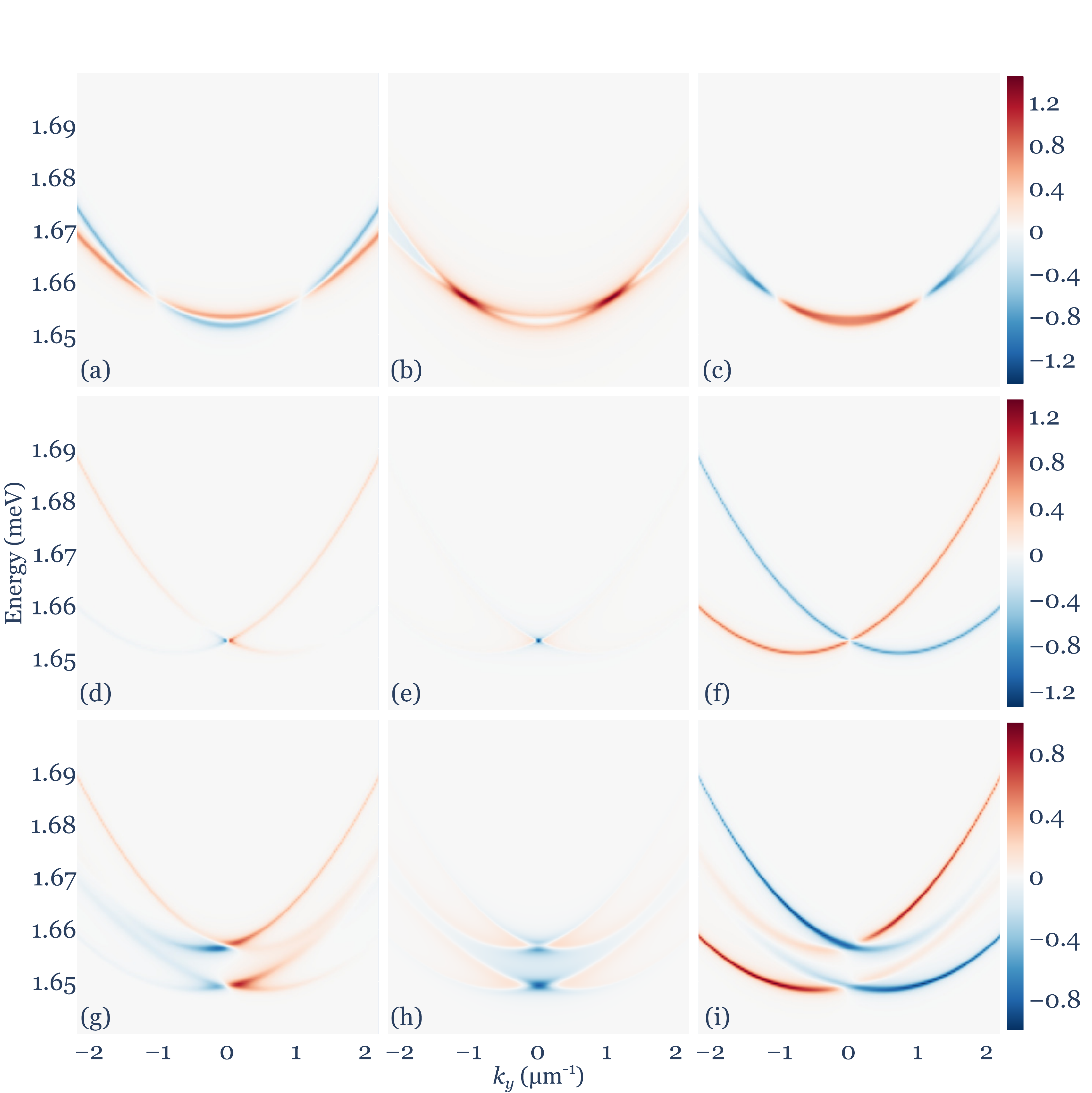}
\caption{\label{fig_s3}\textbf{Stokes parameters of the transmitted cavity light depending on $k_y$ and energy.} The incident light is diagonally polarized. Calculations were performed using the Berreman method at $k_x=0$. Panels from left to right correspond to $S_1$, $S_2$ and $S_3$ respectively. Top row shows the isolated TE-TM + XY system, the middle row shows the isolated RD system, and the bottom row shows the two coupled systems.}
\end{figure}

The Stokes parameters of the transmitted/reflected cavity light $|\Psi\rangle = (\psi_+, \psi_-)^t$ are written in the circular polarization basis as follows, linear vertical-horizontal projection $S_1 = \text{Re}{(\psi_+ \psi_-^*)}$, linear diagonal-antidiagonal projection $S_2 = -\text{Im}{(\psi_+ \psi_-^*)}$, right-hand and left-hand circular projection $S_3 = (|\psi_+|^2 - |\psi_-|^2)/2$, and total light intensity $S = (|\psi_+|^2 + |\psi_-|^2)/2$. Figures~\ref{fig_s1} and~\ref{fig_s2} show the transmitted Stokes parameter distribution and the electric field distribution in reciprocal space, respectively, where we excite at the energy in the second band from the bottom corresponding to maximum Berry curvature [see Fig.~2(c) in main manuscript] (i.e., at $E=1.6603$ eV). We use reciprocal space to present our calculations as they are performed using the Berreman method. Figure~\ref{fig_s3} correspond to transmitted Stokes parameter distribution depending on $k_y$ and energy. We present the distributions for the case of an isolated TE-TM + XY cavity (see Figs.~\labelcref{fig_s1,fig_s2,fig_s3} upper rows), an isolated Rashba-Dresselhaus cavity (see Figs.~\labelcref{fig_s1,fig_s2,fig_s3} middle rows), and for the two coupled cavities (see Figs.~\labelcref{fig_s1,fig_s2,fig_s3} bottom rows).

The top row in Fig.~\ref{fig_s1} shows the expected polarization pattern of a TE-TM + XY system. Two shifted parabolas of different effective masses whose energies correspond to rotating linear polarization eigenmodes as dictated by the effective magnetic field shown in Fig.~1(c) in the main manuscript. The middle row shows the dominant transmission coming from circularly polarized states [see Fig.~\ref{fig_s1}(f)] corresponding to the two momentum displaced valleys of the Rashba-Dresselhaus Hamiltonian. We point out that additional finite Stokes parameters appearing in Fig.~\ref{fig_s1}(c) and Figs.~\ref{fig_s1}(d,e) are due to both finite reflectivity of the modelled cavity and higher order contributions to the cavity mode mixing (see Ref.~\cite{rechcinska2019engineering}).

In the third row of Fig.~\ref{fig_s1} the two coupled cavities result in a nontrivial polarization texture emerging from all four bands in the system. The different amount of transmission across the bands in both frequency and momentum for our chosen incident diagonally polarized light makes analysis quite challenging. Indeed, the spin degree of freedom is entangled with the cavity degree of freedom through the parameter $J$ in Eq.~(5) which forbids us from analysing the system in the framework of two uncoupled pseudospins. As such, the definition of a band’s polarization through the isomorphic relation between the SU(2) and the SO(3) group is lost. None-the-less, the Berreman calculations allow one to characterize the transmitted polarization of the cavity with great detail which becomes a useful tool when comparing theory with experiment. 

\section{Quantum geometric tensor components and Berry curvature of all bands}
\begin{figure}
\includegraphics[width=180mm]{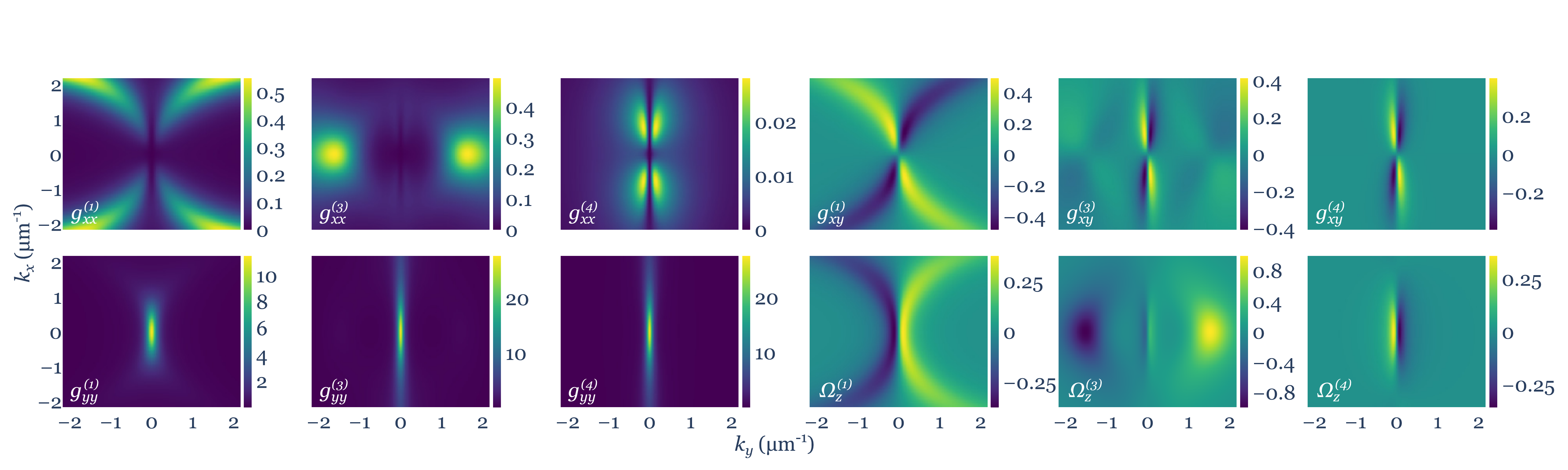}
\caption{\textbf{Quantum geometric tensor components and Berry curvature.} The quantum geometric tensor components $g_{ij}^{(n)}$ and the Berry curvature $\Omega_z^{(n)}$ for bands $n = 1,3,4$ (counted from lowest energy band to highest).}
\label{fig.QGT_allbands}
\end{figure}

Figure~\ref{fig.QGT_allbands} shows the QGT components (see labels in each panel) in the remaining bands $n = 1,3,4$ with band $n=2$ shown in Fig.~2 in the main manuscript.

\providecommand{\noopsort}[1]{}\providecommand{\singleletter}[1]{#1}%


\begin{thebibliography}{45}
\bibitem [{par()}]{parameters}%
  \BibitemOpen
  \href@noop {} {}\bibinfo {note} {Parameters for (a-g) are: $\lambda=750
  \text{ nm}, \ n_{o1} = n_{o2} = 1.5, \ \Delta n_1 = \Delta n_2 = 0.4, \
  m_{x1} = 3, \ m_{y1} = 3 + 4\times 10^{-3} \; (\varphi_1 = 85.18 \text{
  deg}), \ m_{x2} = 7,\ m_{y2} = 8 \ (\varphi_2 = 37.93 \text{ deg})$. In (a)
  and (b) the number of outer DBR layers was $N_1 = 10$ and $N_2 = 12$ with
  dielectric constants $n_{DBR1} = 1.4$ and $n_{DBR2} = 2.3$. For the double
  cavity in (c-h) we used outer layers with $N_1 = N_3 = 10$ and a middle layer
  with $N_2 = 12$. Parameters for (h) are the same except for $m_{y1} = 3 +
  10^{-4} \; (\varphi_1 = 89.24 \text{ deg})$.}\BibitemShut {Stop}%
\end{thebibliography}
\end{document}